\documentclass[final,1p,11pt]{elsarticle}

\usepackage{epsfig}
\usepackage{array,tabularx,epsfig,mathrsfs,graphicx,rotating}
\usepackage{ifthen}
\usepackage{amsfonts}
\usepackage{ragged2e}
\PassOptionsToPackage{hyphens}{url}
\usepackage[hyphens]{url}
\usepackage{hyperref}

\hypersetup{
  colorlinks=true,
  linkcolor=blue,
  citecolor=blue,
  urlcolor=blue
}

\usepackage{breakurl}

\newcommand{\beq}{\begin{equation}}
\newcommand{\eeq}{\end{equation}}

\chardef\til=126

\begin{document}
\begin{frontmatter}

\title{
HepSim: a repository with predictions for high-energy physics experiments
}

\author[lab1]{S.V.~Chekanov\corref{cor1}}
\ead{chekanov@anl.gov}
\cortext[cor1]{Corresponding author}
\address{
HEP Division, Argonne National Laboratory,
9700 S.Cass Avenue,
Argonne, IL 60439, USA. 
}


\begin{abstract}
A file repository for calculations of cross sections and kinematic distributions 
using Monte Carlo generators for high-energy collisions is discussed. The repository 
is used to facilitate effective preservation and archiving of data from theoretical calculations, as well as for
comparisons with experimental data.
The HepSim data library is publicly accessible and  includes a number of Monte Carlo event samples with Standard Model
predictions for current and future experiments.
The HepSim project includes a software package to automate the process of 
downloading and viewing online Monte Carlo event samples. A data streaming over a network for end-user  analysis is discussed.
\end{abstract}


\end{frontmatter}


\section{Introduction}

Modern theoretical predictions quickly become CPU intensive. 
A possible solution to facilitate comparisons between theory and data 
from high-energy physics (HEP) experiments is to develop a public library that stores theoretical predictions
in a form that is suited for calculation of arbitrary experimental distribution on  commodity computers.   
The need for such library is driven by the following modern developments:

\begin{itemize}

\item
The Standard Model (SM) predictions should be substantially improved in order to find
new physics that can potentially exhibit itself within theoretical uncertainties, 
which are currently at the level of 5\% - 10\% for quantum chromodynamics (QCD) theory.
Currently, such uncertainties are the main limiting factor  for precision measurements, as well as 
for searches new physics beyond the SM.  
An increase in theoretical precision  leads to  highly complex, CPU intensive, computations. 
Such calculations are difficult to achieve on commodity computers.
In many cases, it is easier to read events with predictions generated after a proper validation, rather than generating them
for every measurement or experiment.

\item
Searches for new physics often include event scans in different kinematic domains.
This means that the outputs from theoretical predictions should be sufficiently 
flexible  to accommodate  large variations in event selection requirements
and to narrow down  search results. 
A theory ``frozen" in the form of histograms is often difficult to
deal with since histograms need to be computed for each experimental cut.

\item
The current method to generate predictions for experimental papers lacks transparency.
Usually, such calculations are done by experiments using 
computational resources that are often unavailable for theorists.
Theoretical calculations are typically done through  a ``private" communication between data analyzers and theorists, without public access
to the original code or data that are the result of the  computations performed for publications.
For example, common samples with  SM predictions can be useful for a comparison between different experiments 
that often use different selection cuts. 

\end{itemize}

Let us give an  example illustrating the first point. A single calculation of $\gamma$+jet cross section at a next-to-leading order (NLO) QCD
typically requires several hours on a commodity computer.
Sufficient statistical precision for 
a falling transverse momentum spectrum ($p_T(\gamma)$), typical for HEP,  requires 
several independent calculations with different minimum $p_T(\gamma)$ cuts.
Next, the calculations of theoretical uncertainties, such as those with renormalisation scale variations or with 
different sets of the input parton-density 
functions (PDFs), require several additional runs. Thus, a single high-quality prediction for a  publication 
may require up to 10000 CPU hours. Finding a method to store Monte Carlo (MC) events with full systematic variations in  
highly compressed archive files that can be processed by experimentalists and theorists
becomes essential. We will come back to this example in the next sections.

A creation of the library with common data from theoretical models for HEP experiments can be an important step to 
simplify data analysis,  to ensure proper validation, accessibility and preservation over 
the long term for new uses.   
The idea of storing MC predictions (including NLO calculations) in a form of ``n-tuples", i.e. an ordered list of 
records with detailed information on separate (weighted or un-weighted) 
events is not new; one way or the other, many Monte Carlo (MC) and NLO programs
can write data on event-by-event bases into files that can be subsequently read by analysis programs.
The missing part of this approach is a common standard layout for such files, a transparent public access, and  
an easy-to-use software toolkit to process such data for an arbitrary experimental observable. 
The HepSim project aims to achieve this goal.  
 
A number of community projects exist that simplify theoretical computations and comparisons with experimental data, such as 
{\sc MCDB} (a MonteCarlo Database)  \cite{Belov:2007qg}, 
{\sc Professor}  \cite{Abreu:1996na} (a tuning tool for MC event generators), {\sc Rivet} \cite{Buckley:2010ar}
(a toolkit for validation of MC event generators) and {\sc APPLgrid} \cite{applgrid} (a method to reproduce the results of full NLO calculations with any input parton distribution set). In the past, {\sc JetWeb} \cite{Butterworth:2002ts} 
(a WWW interface and database for Monte Carlo tuning) addressed similar questions of comparing data with theory.
Among these tools, the closest repository that  focuses on storing data with theoretical predictions is the {\sc MCDB} Monte Carlo database developed 
within the CMS Collaboration. This publicly available repository mainly includes the {\sc CompHEP} MC events \cite{Pukhov:1999gg}
in the HepML format \cite{Belov:2010xm}. 
 
This paper discusses a public repository with Monte Carlo simulations (including NLO calculations) designed 
for fast calculation of cross sections or any kinematic distribution.
This repository was created during the Snowmass Community Studies \cite{snowmass} in 2013,
that had one of the goals of archiving MC simulation files for future experiments.
In comparison with  the {\sc MCDB} repository, the proposed repository stores 
files in a highly-compressed format that is better suited for archiving, has a simplified data access model with 
a possibility of data streaming from the web, and includes tools to perform calculations of kinematic distributions. 
 
\section{Technical requirements}

A number of software requirements must be met in order 
to achieve the goal of creating an archive of events from theory predictions 
for the HEP community: 

\begin{itemize}

\item
Data should be stored in compact files suitable for network communication. 
In particular, the data format should minimize the usage of fixed-length data types and utilize the "varint" approach
which use fewer bytes for smaller numbers compared to larger, less common,  numbers.
For example, such data serialization is implemented for integer values in the Google's Protocol Buffers library \cite{protobuf}. 
For typical HEP events, large numbers (such as energies, masses, particle identification numbers etc.) are usually less common,
and this can be used for very effective compression.
It is desirable if MC event samples have file sizes of the order of tens of GBs or less for effective exchange and wide usage.

\item
An important requirement for the public access is to be able to read the data in a number
of programming languages, on any computational platform, with a minimum overhead of installing and configuring the
software needed for analysis. 
Therefore, the data format  should be multiplatform from the ground,
with the possibility to process such data on Linux, Windows and Mac computers. Likewise, 
the files should be self-describing and well suited  for structured data, similar to XML. 
The self-describing feature is needed to store data from different MC generators created by different authors,
thus data attributes can be vastly different and should be accessed by name. The documentation of data layout
should be the part of the file, without external documentation of
position field. 
The programming language used to read the data 
should be well suited for concurrency (multi-threading).

\item
Public access via the HTTP protocol is one of the important requirements since this will allow  streaming the simulated 
data to the Web browsers which, in future, can have a functionality of processing and analyzing the data.
Although the samples can be located on the grid, our previous experience shows that sharing event  samples
using the grid access model is less suited for wide community  due to security restrictions. 
A more effective data access, such as the GridFTP protocol,  can be added in future.

\item
When possible, theoretical uncertainties should be encapsulated inside the files. 
For example,
events should include central weights plus all associated systematic variations.
Such ``all in one" approach will significantly simplify the calculations:
A single pass over the data files can be sufficient to create final predictions with all uncertainties. 

\item
In addition to the general availability of the data with simulations, the project should provide  benchmark cross
sections and most representative figures with distributions. All produced plots should be accompanied by analysis programs 
in order to illustrate the data access.

\end{itemize}
The above requirements represent a number of software challenges. For example, the usage 
of the ROOT \cite{root} data-analysis program 
may be insufficient due to (a) ineffective  fixed-length data representation leading to large file sizes.
The usage of the variable-byte encoding leads to files that are $30-40\%$ smaller compared to ROOT and other existing 
fixed-length data formats after compression;  
(b) a complexity in dealing with the C++ system programming language.
From the other hand, the usage of ROOT should be well supported since 
this is the main analysis environment for HEP experiments. 

The choice of the programming language may look obvious at first given that C++
is the preferred  choice of  HEP experiments. However, this can introduce certain limitations 
since C++ requires professional programming expertise that is typically available only for system programmers. 
A scripting languages, such as Python,  should be an essential part of the project.

The scope of this project and its
implementation substantially depend on the usage of high-performance computers.

\section{Current implementation}
\label{imp}

The HepSim repository 
with data samples from leading-order and NLO MC generators is
currently available for validations and checks.
The database is accessible using the URL link given in \cite{hepsim}.

The HepSim project includes the following parts:

\begin{itemize}
  \item A front-end of HepSim with user login;  
  \item A back-end server (or servers) that stores HepSim data;  
  \item A software toolkit that allows to access and process data.
\end{itemize}
Below we will discuss these three parts of the  HepSim  project in more detail.

\subsection{HepSim front-end}

The HepSim has a front-end that stores metadata using a 
SQL database engine. The front-end is written in the PHP and JavaScript languages.
The MC files are stored on a separate file storage with a URL access.  
There is no requirements to store data on the same web server; data can be scattered over multiple URL locations and
it is up to the user to document data locations.
The SQL front-end can be used to search the database using dataset description, 
MC generator name, production process,  cross section and other metadata that are included
in the description of MC files. 

\begin{figure}[th]
\begin{center}
\includegraphics[width=0.95\textwidth]{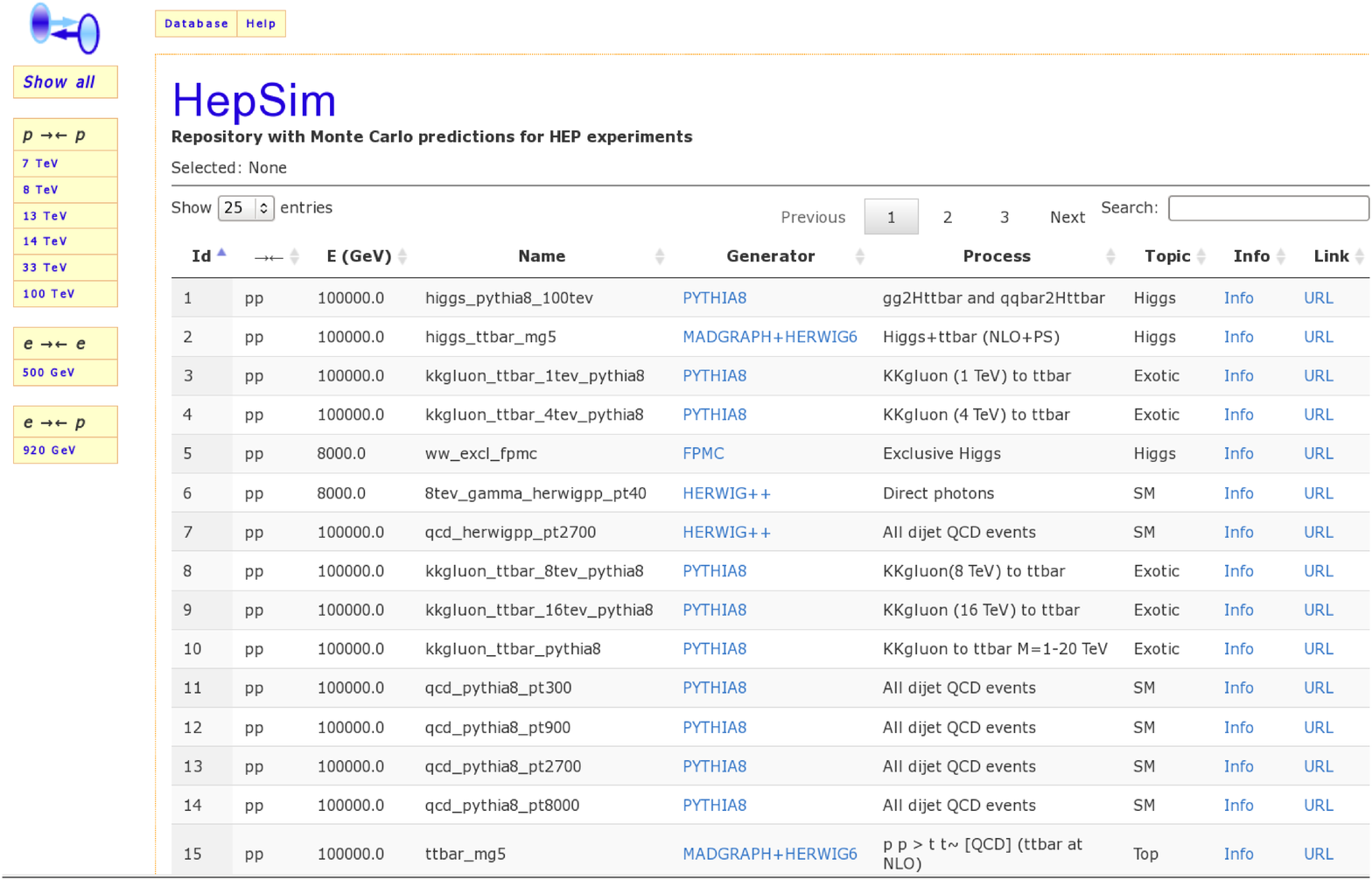}
\end{center}
\caption{
The HepSim front-end to access  MC datasets.
}
\label{fig:hepsim_db}
\end{figure}

In order  to add an entry to the database, a user should be registered.
Upon the registration, a dataset should be added by creating a metadata record with  
dataset name, physics process, the name of the MC generator, file sizes, a text short description, 
file format
and the URL location of the dataset. 
Figure~\ref{fig:hepsim_db} shows the HepSim database front-end that lists available samples.

The help menu of  HepSim  describes how to perform a bulk download of multiple files from the repository and 
how to read events using minimum requirements for software setup. A more advanced usage is explained on the  wiki
linked to the HepSim web page.

\subsection{HepSim back-end}

The front-end of HepSim includes URL links to the actual data that are located on a separate data server. 
The HepSim data servers can be distributed in several locations, since the front-end does not impose
any particular requirements on the data servers.

As a basis for the HepSim public library,
the ProMC \cite{2013arXiv1306.6675C,2013arXiv1311.1229C} file format  has been chosen.
This choice is motivated  by the possibility  
to store data with arbitrary layout using variable-byte encoding, including 
log files from MC generators, and the possibility to stream data over the network. 
ProMC is implemented as a simple, self-containing library that can easily be deployed on a
number of platforms including high-performance computers, such as IBM BlueGene/Q.

The ProMC format is based on  a dynamic assignment of the needed number of bytes
to store integer values, unlike the traditional approaches that use 
the fixed-length byte representations. 
The advantage of this ``varint`` feature has been discussed in  \cite{2013arXiv1306.6675C,2013arXiv1311.1229C}. 
We will illustrate this using another example relevant to NLO QCD calculations.
To store a single event together with theoretical uncertainties created by a NLO program, one needs
to write the information on a few particles together with the event weights representing theoretical uncertainties. 
For the $\gamma$+jet example discussed previously, we need to store a few  particles from the hard scattering
(where one outgoing particle is photon), together with event weights from  different sets of  PDFs. 
Although the central weight can be stored as a floating point number 
without losing the numerical precision, other weights can be encoded as integer values 
representing deviations from the central weight. This approach can take the advantage  
of the compact varint ``compression''.
For example, if a central weight, denoted as $PDF(0)$, is estimated with  MSTW2008 PDF \cite{Martin:2009iq},
40 associated eigenvector sets for PDF uncertainties can be represented as integer numbers: 
$$
w_n = \Big[ 1000\times (1- \frac{PDF(n)}{PDF(0)}) \Big], \qquad n=1,..,40, 
$$ 
i.e. in the units of $0.1\%$ with respect to the central wight $PDF(0)$.  The factor 1000 is arbitrary and can be 
changed depending on the required precision. In many cases, integer values 
$w_n$ are close to 0, leading to 1-2 bytes in the varint encoding.
Therefore, a single event record with all associated eigenvector PDF sets will use less than 100 bytes. 

The ProMC files can be  read in 
a number of programming languages supported on the Linux and Windows platforms. 
The default language to read and process files for validation purposes was chosen to be Java, since it is 
well suited for web-application programming and is available on all major computational platforms.
To process data from HepSim, the ROOT data-analysis program \cite{root} developed at CERN
can be used. In addition to the ProMC, the HepSim 
database can include datasets in other popular formats, such as HepML \cite{Belov:2010xm},
HEPMC \cite{hepmc} or StdHEP \cite{stdhep}.

\subsection{Available datasets}

Currently, the  HepSim repository contains events generated by  {\sc PYTHIA}~\cite{pythia}, {\sc MadGraph}~\cite{Alwall:2011uj}, 
{\sc Jetphox}~\cite{jetphox1}, {\sc MCFM}~\cite{Campbell:2010ff}, {\sc NLOJet++}~\cite{Nagy:2003tz}, FPMC~\cite{Boonekamp:2011ky}  and {\sc HERWIG++}~\cite{Herwig} generators.
The repository includes  event samples for $pp$ colliders with the centre-of-mass energies of 8, 13, 14, 100~TeV.
In some cases,  together with the detailed information on produced particles, 
full sets of theoretical uncertainties (scale, PDF, etc.) are embedded inside the files as discussed in Sect.~\ref{imp}.
A number of simulated samples were created using  the IBM BlueGene/Q (located at the
Argonne Leadership Computing Facility) and the ATLAS Connect virtual cluster service, 
the descriptions of which are beyond the scope of this paper.

The total size of a typical HepSim  dataset  
is less than 100 GB.  
The largest simulated sample stored in HepSim  and used in physics studies \cite{Auerbach:2014xua} 
contains 400  million $pp$ collision events at the center-of-mass energy
of $100$~TeV.  
Each event contains more than 5000 particles on average, totaling more than 2 trillion generated particles. The total size of this sample is 4.2~TB. 

Each particle in a typical HepSim  dataset is characterized by four-momentum, position and several quantum numbers. 
In many cases, the event records are ``slimmed'' after removing unstable particles and final-state particles
with transverse momentum less than 300--400~MeV. The most essential parton-level information on vector bosons, $b-$ and $t-$ quarks
is kept. 

When possible, ProMC files with NLO predictions include deviations
from the central event weight in the form of integer values as discussed in the previous section. 
This typically leads to a very compact representation of events from NLO generators using the ``varint'' encoding 
since large  systematic deviations are less common than small ones.

\section{HepSim software toolkit}

The HepSim toolkit is designed for download, validation and viewing Monte Carlo event samples.   
On Linux/Mac, the HepSim software can be downloaded and installed as:

\begin{verbatim}
curl http://atlaswww.hep.anl.gov/asc/hepsim/hs-toolkit.tgz | tar -xz 
source hs-toolkit/setup.sh
\end{verbatim}
To use this package, Java 7 (or 8) should be installed. 
There is no other requirements to use this package. 

Let us consider several commands from the package ``hs-toolkit'' 
that can help to download and analyze HepSim Monte Carlo samples.

\begin{itemize}

\item
The command to show all files associated with a given dataset is: 

\begin{verbatim}
hs-ls [name]
\end{verbatim}
where ``[name]''  is the dataset name. Alternatively, the dataset name can be replaced
with the URL location of the dataset on the web. 

\item
To search for a specific URL by name or dataset description, use this command: 
\begin{verbatim}
hs-find [word]
\end{verbatim}
where ``[word]'' is a word that matches your criteria. This command returns a list of sites where the given word is present in the dataset names or description. 

\item
The files can be downloaded in a multiple threads as:
\begin{verbatim}
hs-get [name] [OUTPUT DIR] [Nr of threads] [Nr of files] [Pattern] 
\end{verbatim}
where ``[OUTPUT DIR]'' is the name of the output directory, ``[Nr of threads]'' is 
the number of threads for data download, ``[Nr of files]'' is the maximum number of files for download
and [Pattern] is 
an (optional)  pattern that the regular expression engine attempts to match in the file names.
Alternatively, the dataset name ``[name]'' can be replaced  
with the URL of the dataset.                              

\item
To check a single file from the dataset and to print its metadata, one can use the following command:
\begin{verbatim}
hs-info [URL] 
\end{verbatim}
Note that ``[URL]'' can either be a file URL or an absolute path of the file on a disk.
This command is slower in the case of URL.
In order to print an event on a Linux/Mac console, use:
\begin{verbatim}
hs-info [URL] [Event number] 
\end{verbatim}
where the last argument is an event (integer) number.

\item
In order to look at all events using a GUI mode, use this command:

\begin{verbatim}
hs-view [URL] 
\end{verbatim}
where, again,  ``[URL]'' can  either be a URL or the location of a file on the local disk.
There is no limitation on file sizes for this command.
\end{itemize}

The above examples illustrate the fact that data can be streamed over a network, without storing data on the disk.
The only limitation for this approach is the computer memory.  

\subsection{Data validation}

For validations of HepSim simulated samples,
Jython, an implementation of the Python programming language in Java, is used. 
The Jython language has similar semantics to Python, but uses the Java Virtual Machine (JVM) which
ensures platform independence of the analysis environment.
The Jython scripts can straightforwardly be rewritten in  Groovy, JRuby and other
scripting  dynamic languages supported by JVM.   
In many cases, the validation code examples accompany the data sets and are publicly available for the users. 
The validation codes  show how to read the ProMC files with simulation data and how
to reconstruct cross sections when the event weights are required (i.e. for NLO programs).
The Jython snippets were written using the SCaVis~\cite{scavis,Chekanov:1261772} data-analysis framework for the Java platform, 
but any Java-based IDE (Eclipse, NetBeans or IntelliJ) should be  sufficient to develop the codes 
as long as the needed jar libraries are included in the Java classpath.
All basic analysis packages for HEP physics, such as a four-vector with the Lorentz transformations and
different types of jet reconstruction algorithms, such as  the popular $k_T$, anti-$k_T$ and Cambridge/Aachen inclusive jet 
clustering algorithms (\cite{Ellis:1993tq,Cacciari:2008gp} and references therein)
for $pp$ collisions,
are supported by the SCaVis Java libraries.
Jet algorithms for $e^+e^-$ collisions are supported via the FreeHEP Java library \cite{freehep}.

\begin{figure}[th]
\begin{center}
\includegraphics[width=0.5\textwidth]{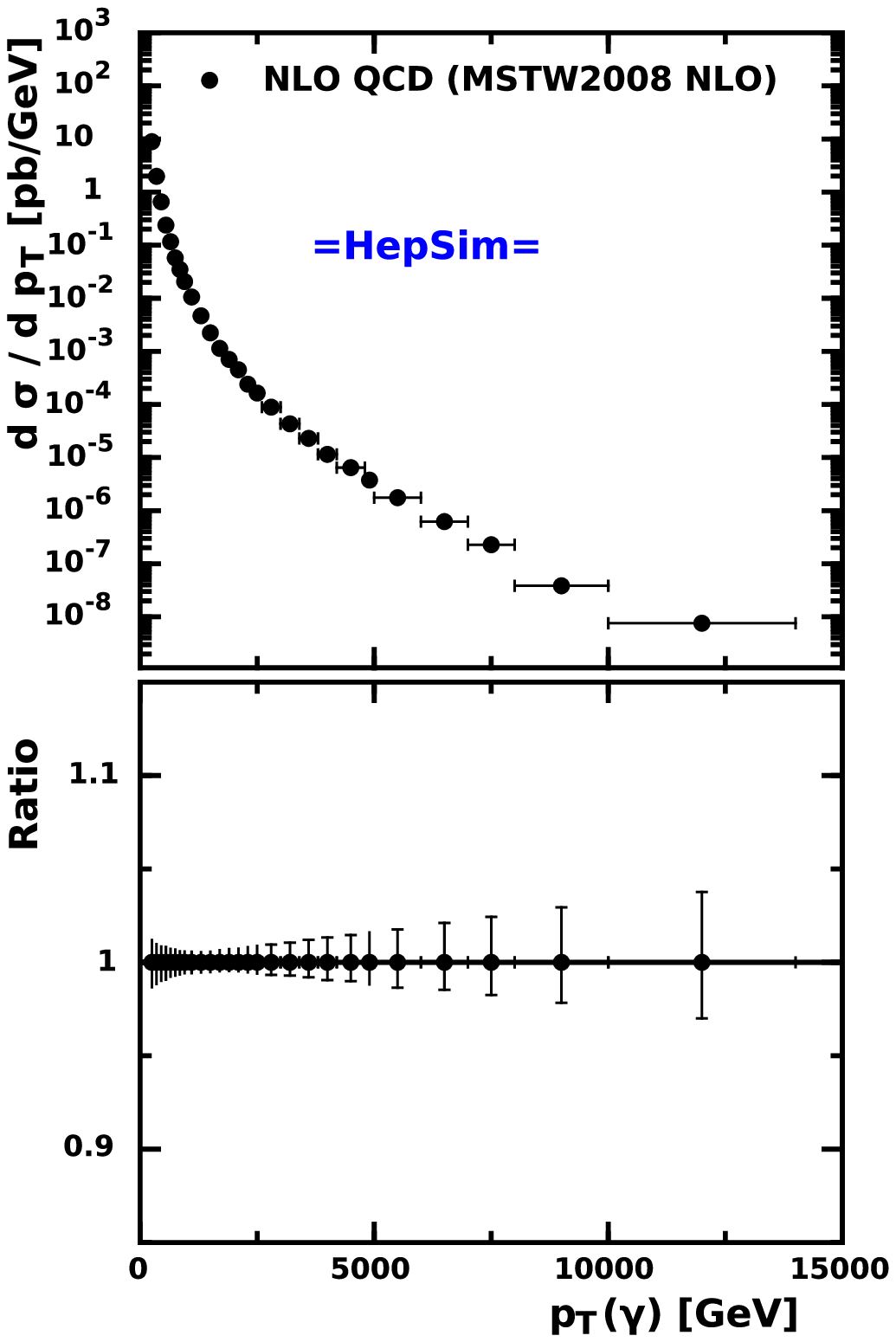}
\end{center}
\caption{
A typical example of the HepSim library output showing the {\sc Jetphox} NLO \cite{jetphox1,PhysRevD.73.094007}
calculation for $\gamma$+jet cross section for a 100 TeV
$pp$ collider.  
The shaded band on the bottom plot shows the relative theoretical uncertainty estimated from 41
PDF sets of the MSTW2008 NLO PDF \cite{Martin:2009iq}.
This validation plot was created using a Jython code on the Java platform.
}
\label{fig:graph}
\end{figure}

The validation scripts can read data either through the HTTP protocol, or using files stored on local
file systems.  The analysis of files using streaming over a network is typically slower, and thus 
is only recommended for a non-repeatable  analysis.

The processing validation time for a typical simulated sample is less than 30~min on a desktop computer while, 
in some cases, the CPU time to generate such event samples is more than 8000 CPU hours (512 nodes times 16 cores) 
on IBM BlueGene/Q of the  
Argonne Leadership Computing Facility. 

As an example, 
Fig.~\ref{fig:graph} shows the $\gamma$+jet differential cross section for a $pp$ collider
at a center-of-mass energy of 100~TeV created using a validation script.  
The {\tt JETPHOX}~1.3 program~\cite{jetphox1,PhysRevD.73.094007}, which implements a full
NLO QCD calculation of both the direct and fragmentation contributions
to the total cross section, was used to generate the prediction. 
The bottom plot shows the PDF uncertainty calculated from $N=41$ PDF weights provided by the MSTW2008 NLO PDF \cite{Martin:2009iq}:
$$
\frac{\sqrt{\sum_{i=1}^{N} (\sigma_i - \sigma_0)^2}}{\sigma_0}, 
$$
where $\sigma_i$ is the cross section for the $i^{th}$ eigenvector of the MSTW2008 NLO set,
and $\sigma_0$ is the cross section for the central MSTW2008 NLO set. Negative and positive
values of the difference $(\sigma_i - \sigma_0)$  are treated separately. 
The output data sample is about 5~GB and includes 7 calculations generated with different minimum $p_T(\gamma)$ cuts.
All PDF weights are included in the file record using the variable-byte encoding discussed before.
The processing time is 30~min on a commodity computer using a Jython script which
reads 4-momenta of particles and event weights. Thus, any distribution with arbitrary experimental cuts and histogram bin sizes
can be repeated within this time.

\subsection{Data analysis}

As discussed before, the  HepSim repository is useful for a fast reconstruction of theoretical cross
sections and distributions from four-momenta of particles using experiment-specific selection,
reconstruction and histogram bins.
The analysis code in C++, Java and Python can be generated from 
the ProMC files as described in Ref.~\cite{2013arXiv1306.6675C,2013arXiv1311.1229C} 

For a full-scale analysis of HepSim data samples, the ROOT data-analysis program \cite{root} developed at CERN
can be used.  How to compile data-analysis programs with ROOT is given in a number of examples that come with the
ProMC package (inside the directory ``examples''). The analysis can also be done using ROOT I/O, after converting
data to the ROOT file format.  However, this might be redundant since data in the ProMC format can be read by C++ programs
directly.              

The HepSim files can also be used as inputs for the {\sc DELPHES}
fast detector simulation program  \cite{deFavereau:2013fsa} which has
a built-in reader for ProMC files.

\section{Summary}

The online HepSim manual \cite{hepsim} contains a description of how to search for simulated samples, download them in multiple threads,  
how to read data using Java, C++/ROOT and CPython and how to run a fast detector simulation.
Currently, the database includes more than 70 event samples that cover a wide range of physics processes for 
$pp$ collision energies from 7~TeV to 100~TeV. 
A number of publications  based on the HepSim database are listed on the  web page. 
The current focus of HepSim is to publish simulated events
for the High Luminosity LHC and for studies of the physics potential of a future 100 TeV $pp$ collider.

\section*{Acknowledgments}
I would like to thank J.~Proudfoot and E.~May for discussion and validation. 
The submitted manuscript has been created by UChicago Argonne, LLC,
Operator of Argonne National Laboratory (``Argonne'').
Argonne, a U.S. Department of Energy Office of Science laboratory,
is operated under Contract No. DE-AC02-06CH11357.
This research used resources of the Argonne Leadership Computing Facility at Argonne National Laboratory, which is supported by the Office of Science of the U.S. Department of Energy under contract DE-AC02-06CH11357. A fraction of the simulated event samples presented in this paper  were
generated using the ATLAS Connect virtual cluster service.

\section*{References}

\bibliographystyle{elsarticle-num}
\def\bibname{\Large\bf References}
\def\refname{\Large\bf References}
\pagestyle{plain}
\bibliography{biblio}

\end{document}